\documentstyle[12pt]{article}
\newcommand{\cl}{\centerline}
\newcommand{\beq}{\begin{equation}}
\newcommand{\eeq}{\end{equation}}
\newcommand{\bea}{\begin{eqnarray}}
\newcommand{\eea}{\end{eqnarray}}

\begin{document}
\begin{titlepage}
\setlength{\textwidth}{5.0in} 
\setlength{\textheight}{7.5in}
\setlength{\parskip}{0.0in} 
\setlength{\baselineskip}{18.2pt}
\hfill {\tt SOGANG-HEP 275/00}
\begin{center}
\cl{\Large{{\bf Flavor symmetry breaking effects}}}\par
\cl{\Large{{\bf on SU(3) Skyrmion}}}\par 
\vskip 0.5cm
\begin{center}
{Soon-Tae Hong$^a$ and Young-Jai Park$^{b}$}\par
\end{center}
\vskip 0.4cm
\begin{center}
{Department of Physics and Basic Science Research Institute,}\par
{Sogang University, C.P.O. Box 1142, Seoul 100-611, Korea}\par
\end{center}
\vskip 0.5cm
\cl{\today}
\vfill
\begin{center}
{\bf ABSTRACT}
\end{center}
\begin{quotation}
We study the massive SU(3) Skyrmion model to investigate the flavor symmetry 
breaking (FSB) effects on the static properties of the strange baryons in the 
framework of the rigid rotator quantization scheme combined with the improved 
Dirac quantization one.  Both the chiral symmetry breaking pion mass and FSB 
kinetic terms are shown to improve $c$ the ratio of the strange-light to 
light-light interaction strengths and $\bar{c}$ that of the strange-strange 
to light-light. 

\par
---------------------------------------------------------------------\\
\noindent $^a$sthong@ccs.sogang.ac.kr\\ 
\noindent $^b$yjpark@ccs.sogang.ac.kr 
\vskip 0.5cm 
\noindent
\end{quotation}
\end{center}
\end{titlepage}

It is well known that baryons can be obtained from topological solutions,
known as SU(2) Skyrmions, since homotopy group $\Pi_{3}(SU(2))=Z$ admits
fermions~\cite{adkins83,sk,hsk}. Using collective coordinates of isospin
rotation of the Skyrmion, Adkins et al. \cite{adkins83} have performed
semiclassical quantization having static properties of baryons within 30\%
of the corresponding experimental data. The hyperfine splittings for the
SU(3) Skyrmion~\cite{su3} has been studied in two main schemes. Firstly, the 
SU(3) cranking method exploits rigid rotation of the Skyrmion in the collective
space of SU(3) Euler angles with full diagonalization of the flavor symmetry
breaking (FSB) terms~\cite{fsbsu3}. Especially, Yabu and Ando~\cite{yabu} 
proposed the exact diagonalization of the symmetry breaking terms by 
introducing higher irreducible representation mixing in the baryon wave 
function, which was later interpreted in terms of the multiquark 
structure~\cite{multi} in the baryon wave function. Secondly, Callan and 
Klebanov~\cite{callan} suggested an interpretation of baryons containing a 
heavy quark as bound states of solitons of the pion chiral Lagrangian with 
mesons. In their formalism, the fluctuations in the strangeness direction are 
treated differently from those in the isospin directions~\cite{callan,sco}.

On the other hand, the Dirac method~\cite{di} is a well known formalism to
quantize physical systems with constraints. In this method, the Poisson
brackets in a second-class constraint system are converted into Dirac
brackets to attain self-consistency. The Dirac brackets, however, are
generically field-dependent, nonlocal and contain problems related to
ordering of field operators. These features are unfavorable for finding
canonically conjugate pairs.  To overcome the above problems, Batalin, 
Fradkin, and Tyutin (BFT)~\cite{BFT} developed a method which converts the 
second-class constraints into first-class ones by introducing auxiliary 
fields. Recently, this BFT scheme has been successively applied to several 
models of current interest~\cite{BFT1,kpr}.  Especially this BFT 
method~\cite{BFT} has given an additional energy term in SU(2) Skyrmion 
model~\cite{hkp98} and has been also applied~\cite{hong00} to an open string 
theory with $D$-branes.

The motivation of this paper is to generalize the standard flavor symmetric 
(FS) SU(3) Skyrmion rigid rotator approach~\cite{kleb94} to the SU(3) Skyrmion
 case with the pion mass and FSB terms so that one can investigate the chiral 
breaking pion mass and FSB effects on $c$ the ratio of the strange-light to 
light-light interaction strengths and $\bar{c}$ that of the strange-strange to 
light-light. 

Now we start with the SU(3) Skyrmion Lagrangian of the form 
\begin{eqnarray}
{\cal L}&=&-\frac{1}{4}f_{\pi}^{2}{\rm tr}(l_{\mu}l^{\mu}) +\frac{1}{32e^{2}}%
{\rm tr}[l_{\mu},l_{\nu}]^{2}+{\cal L}_{WZW}  \nonumber \\
& &+\frac{1}{4}f_{\pi}^{2}{\rm tr}M(U+U^{\dag}-2)+{\cal L}_{FSB},  \nonumber
\\
{\cal L}_{FSB}&=&\frac{1}{6}(f_{K}^{2}m_{K}^{2}-f_{\pi}^{2}m_{\pi}^{2}) {\rm %
tr}((1-\sqrt{3}\lambda_{8})(U+U^{\dag}-2))  \nonumber \\
& &-\frac{1}{12}(f_{K}^{2}-f_{\pi}^{2}){\rm tr} ((1-\sqrt{3}%
\lambda_{8})(Ul_{\mu}l^{\mu} +l_{\mu}l^{\mu}U^{\dag})),
\end{eqnarray}
where $f_{\pi}$ and $f_{K}$ are the pion and kaon decay constants. Here $e$
is the dimensionless Skyrme parameter and $l_{\mu}=U^{\dag}\partial_{\mu}U$ with an
SU(3) matrix $U$ and $M$ is proportional to the quark mass matrix given by 
\[
M={\rm diag}~ (m_{\pi}^{2},~ m_{\pi}^{2},~ 2m_{K}^{2}-m_{\pi}^{2}), 
\]
where $m_{\pi}=138$ MeV and $m_{K}=495$ MeV. Note that ${\cal L}_{FSB}$ is
the FSB correction term due to the relations $m_{\pi}\neq m_{K}$ and 
$f_{\pi}\neq f_{K}$~\cite{pari91,hong93} and the Wess-Zumino-Witten (WZW) term~\cite{wzw} is 
described by the action 
\[
\Gamma_{WZW}=-\frac{iN}{240\pi^{2}}\int_{{\sf M}}{\rm d}^{5}r\epsilon^{\mu%
\nu \alpha\beta\gamma}{\rm tr}(l_{\mu}l_{\nu}l_{\alpha}l_{\beta}l_{\gamma}), 
\]
where $N$ is the number of colors and the integral is done on the
five-dimensional manifold ${\sf M}=V\times S^{1}\times I$ with the
three-space volume $V$, the compactified time $S^{1}$ and the unit interval $%
I$ needed for a local form of WZW term.

Now we consider only the rigid motions of the SU(3) Skyrmion 
\[
U(\vec{x},t)={\cal A}(t)U_{0}(\vec{x}){\cal A}(t)^{\dag}. 
\]
Assuming maximal symmetry in the Skyrmion, we describe the hedgehog solution 
$U_{0}$ embedded in the SU(2) isospin subgroup of SU(3) 
\[
U_{0}(\vec{x})=\left( 
\begin{array}{cc}
e^{i\vec{\tau}\cdot\hat{x}f(r)} & 0 \\ 
0 & 1
\end{array}
\right), \label{u} 
\]
where the $\tau_{i}$ ($i$=1,2,3) are Pauli matrices, $\hat{x}=\vec{x}/r$ and 
$f(r)$ is the chiral angle determined by minimizing the static mass $E$
given below and for unit winding number $\lim_{r \rightarrow \infty} f(r)=0$
and $f(0)=\pi$.

Since ${\cal A}$ belongs to $SU(3)$, ${\cal A}^{\dag}\dot{{\cal A}}$ is
anti-Hermitian and traceless to be expressed as a linear combination of $%
i\lambda_{a}$ as follows 
\[
{\cal A}^{\dag}\dot{{\cal A}}=ief_{\pi}v^{a}\lambda_{a}=ief_{\pi}\left( 
\begin{array}{cc}
\vec{v}\cdot\tau +\nu 1 & V \\ 
V^{\dag} & -2\nu
\end{array}
\right) 
\]
where 
\begin{equation}
\vec{v}=(v^{1},v^{2},v^{3}),~~V=\left( 
\begin{array}{c}
v^{4}-iv^{5} \\ 
v^{6}-iv^{7}
\end{array}
\right),~~ \nu=\frac{v^{8}}{\sqrt{3}}.  \label{vs}
\end{equation}
After tedious algebraic manipulations, the FSB contribution to the Skyrmion
Lagrangian is then expressed as 
\begin{eqnarray}
{\cal L}_{FSB}&=&-(f_{K}^{2}m_{K}^{2}-f_{\pi}^{2}m_{\pi}^{2})(1-\cos
f)\sin^{2}d  \nonumber \\
& &+\frac12 (f_{K}^{2}-f_{\pi}^{2})\sin^{2}d \left(\frac{8}{3}%
e^{2}f_{\pi}^{2}\vec{v}^{2}\sin^{2}f-\frac{2\sin^{2}f}{r^{2}} -\left(\frac{%
{\rm d}f}{{\rm d}r}\right)^{2}\right)\cos f  \nonumber \\
& &-(f_{K}^{2}-f_{\pi}^{2})e^{2}f_{\pi}^{2}\frac{\sin^{2}d}{d^{2}}
\left((1-\cos f)^{2}\| D^{\dag}V\|^{2}-\sin^{2}f\| D^{\dag} \tau\cdot\hat{r}%
V\|^{2}\right)  \nonumber \\
& &+\frac{i\sqrt{2}}{3}(f_{K}^{2}-f_{\pi}^{2})e^{2}f_{\pi}^{2}\frac{\sin 2d}{%
d} \sin^{2} f (D^{\dag}\vec{v}\cdot\tau V -(D^{\dag}\vec{v}\cdot\tau V)^{*})
\nonumber \\
& &+(f_{K}^{2}-f_{\pi}^{2})e^{2}f_{\pi}^{2}\cos^{2}d (1-\cos f)V^{\dag}V.
\label{fsblag}
\end{eqnarray}

In order to separate the SU(2) rotations from the deviations into strange
directions, the time-dependent rotations can be written as~\cite{kleb90} 
\[
{\cal A}(t)=\left( 
\begin{array}{cc}
A(t) & 0 \\ 
0 & 1
\end{array}
\right)S(t) 
\]
with $A(t) \in$ SU(2) and the small rigid oscillations $S(t)$ around the
SU(2) rotations.\footnote{Here one notes that the fluctuations $\phi_{a}$ from 
collective rotations $A$ can be also separated by the other suitable 
parametrization~\cite{schw92}
$
U=A\sqrt{U_{0}}A^{\dagger}{\rm exp}(i\sum_{a=1}^{8}\phi_{a}\lambda_{a})
A\sqrt{U_{0}}A^{\dagger}.
$
} Furthermore, we exploit the time-dependent angular velocity of the SU(2) 
rotation through 
\[
A^{\dagger}\dot{A}=\frac{i}{2}\dot{\alpha}\cdot\vec{\tau}.
\]
Note that one can use the Euler angles for the parameterization of the rotation
~\cite{schwei91}.  On the other hand the small rigid oscillations $S$, which were also used in Ref.~\cite{kleb94}, can be described as 
\[
S(t)={\rm exp}(i\sum_{a=4}^{7}d^{a}\lambda_{a})={\rm exp}(i{\cal D}), 
\]
where 
\[
{\cal D}=\left( 
\begin{array}{cc}
0 & \sqrt{2}D \\ 
\sqrt{2}D^{\dag} & 0
\end{array}
\right),~~ D=\frac{1}{\sqrt{2}}\left( 
\begin{array}{c}
d^{4}-id^{5} \\ 
d^{6}-id^{7}
\end{array}
\right). 
\]

Including the FSB correction terms in Eq. (\ref{fsblag}), the Skyrmion
Lagrangian to order $1/N$ is then given in terms of the angular velocity $\alpha_{i}$ and the strange deviations $D$ 
\begin{eqnarray}
L&=&-E+\frac{1}{2}{\cal I}_{1}\dot{\alpha}\cdot\dot{\alpha}+(4{\cal I}_{2}
+\Gamma_{1})\dot{D}^{\dag}\dot{D}+\frac{i}{2}N(D^{\dag}\dot{D} -\dot{D}^{\dag}D) 
\nonumber \\
& &+i({\cal I}_{1}-2{\cal I}_{2}-\frac12 \Gamma_{1}+\Gamma_{2})
\left(D^{\dag}\dot{\alpha}\cdot\vec{\tau}\dot{D}-\dot{D}^{\dag}\dot{\alpha}\cdot\vec{\tau}D\right)
\nonumber\\
& &-\frac{1}{2}ND^{\dag}\dot{\alpha}\cdot\vec{\tau}D  
+2\left({\cal I}_{1}-\frac{4}{3}{\cal I}_{2}-\frac{4}{3}\Gamma_{1}
+3\Gamma_{2}\right)(D^{\dag}D)(\dot{D}^{\dag}\dot{D})  \nonumber \\
& &-\frac{1}{2}\left({\cal I}_{1}-\frac{4}{3}{\cal I}_{2}-\frac{1}{3}
\Gamma_{1}+2\Gamma_{2}\right)(D^{\dag}\dot{D}+\dot{D} ^{\dag}D)^{2} 
\nonumber \\
& &+\left(2{\cal I}_{2}+\frac12 \Gamma_{1}\right)(D^{\dag}\dot{D} -\dot{D}%
^{\dag}D)^{2} -\frac{i}{3}N(D^{\dag}\dot{D}-\dot{D}^{\dag}D)D^{\dag}D 
\nonumber \\
& &-\frac12\Gamma_{0} m_{\pi}^{2}-\left(\Gamma_{0}(\chi^{2}m_{K}^{2}
-m_{\pi}^{2})+\Gamma_{3}\right)\left(D^{\dag}D -\frac{2}{3}%
(D^{\dag}D)^{2}\right)  \nonumber \\
& &-2(\Gamma_{1}-\Gamma_{2})(D^{\dag}\dot{D})(\dot{D}^{\dag}D),  \label{lag}
\end{eqnarray}
where $\chi=f_{K}/f_{\pi}$. Here the soliton energy $E$, the moments of
inertia ${\cal I}_{1}$ and ${\cal I}_{2}$, the strength $\Gamma_{0}$ of the
chiral symmetry breaking and the inertia parameters $\Gamma_{i}$ $(i=1,2,3)$
originated from the FSB term are respectively given by 
\begin{eqnarray}
E&=&4\pi\int_{0}^{\infty}{\rm d}r r^{2}\left[\frac{f_{\pi}^{2}}{2}
\left(\left(\frac{{\rm d}f} {{\rm d}r}\right)^ {2} +\frac{2\sin^{2}f}{r^{2}}%
\right)\right.  \nonumber \\
& &\left.+\frac{1}{2e^{2}} \frac{\sin^{2}f}{r^{2}} \left(2\left(\frac{{\rm d}%
f}{{\rm d}r}\right)^{2} +\frac{\sin^{2}f}{r^{2}}\right)\right],  \nonumber \\
{\cal I}_{1}&=&\frac{8\pi}{3}\int_{0}^{\infty}{\rm d}r r^{2}\sin^{2}f\left[%
f_{\pi}^{2} +\frac{1}{e^{2}}\left(\left( \frac{{\rm d}f}{{\rm d}r}%
\right)^{2}+\frac{\sin^{2}f}{r^{2}}\right)\right],  \nonumber \\
{\cal I}_{2}&=&2\pi\int_{0}^{\infty}{\rm d}r r^{2}(1-\cos f) \left[%
f_{\pi}^{2}+\frac{1}{4e^{2}}\left(\left(\frac{{\rm d}f}{{\rm d}r}\right)^{2}
+\frac{2\sin^{2}f}{r^{2}}\right)\right],  \nonumber \\
\Gamma_{0}&=&8\pi f_{\pi}^{2}\int_{0}^{\infty}{\rm d}r r^{2}(1-\cos f), 
\nonumber \\
\Gamma_{1}&=&(\chi^{2}-1)\Gamma_{0},  \nonumber \\
\Gamma_{2}&=&(\chi^{2}-1)\frac{8\pi}{3}f_{\pi}^{2}\int_{0}^{\infty}{\rm d}r
r^{2}\sin^{2}f,  \nonumber \\
\Gamma_{3}&=&(\chi^{2}-1)4\pi f_{\pi}^{2}\int_{0}^{\infty}{\rm d}r r^{2}
\left(\left(\frac{{\rm d}f}{{\rm d}r}\right)^{2} +\frac{2\sin^{2}f}{r^{2}}
\right)\cos f.  \label{eni}
\end{eqnarray}

The momenta $\pi_{h}^{i}$ and $\pi_{s}^{\alpha}$, conjugate to the collective
coordinates $\alpha_{i}$ and the strange deviation $D_{\alpha}^{\dag}$ are
given by 
\begin{eqnarray}
\vec{\pi}_{h}&=&{\cal I}_{1}\dot{\alpha}+i\left({\cal I}_{1} -2{\cal I}%
_{2}-\frac12\Gamma_{1}+\Gamma_{2}\right)\left(D^{\dag}\vec{\tau}\dot{D}
-\dot{D}^{\dag}\vec{\tau}\right)-\frac{1}{2}ND^{\dag}\vec{\tau}D,  \nonumber \\
\pi_{s}&=&(4{\cal I}_{2}+\Gamma_{1})\dot{D}-\frac{i}{2}ND
-i\left({\cal I}_{1}-2{\cal I}_{2}-\frac12\Gamma_{1}+\Gamma_{2}\right)
\dot{\alpha}\cdot \vec{\tau}D  \nonumber \\
& &+2\left({\cal I}_{1}-\frac{4}{3}{\cal I}_{2}-\frac{4}{3}\Gamma_{1}
+3\Gamma_{2}\right)(D^{\dag}D)\dot{D}  \nonumber \\
& &-\left({\cal I}_{1}-\frac{4}{3}{\cal I}_{2}-\frac{1}{3}\Gamma_{1}
+2\Gamma_{2}\right)(D^{\dag}\dot{D}+\dot{D}^{\dag}D)D  \nonumber \\
& &-(4{\cal I}_{2}+\Gamma_{1})(D^{\dag}\dot{D}-\dot{D}^{\dag}D)D +\frac{i}{3}%
N(D^{\dag}D)D  \nonumber \\
& &-2(\Gamma_{1}-\Gamma_{2})(D^{\dag}\dot{D})D,  \label{conjmoms}
\end{eqnarray}
which satisfy the Poisson brackets 
\[
\{\alpha_{i},\pi_{h}^{j}\}=\delta_{i}^{j},~~~\{D_{\alpha}^{\dag},
\pi_{s}^{\beta}\}=\{D^{\beta},\pi_{s,\alpha}^{\dag}\}=\delta_{\alpha}^{%
\beta}. 
\]
Performing Legendre transformation, we obtain the Hamiltonian to order $1/N$
as follows 
\begin{eqnarray}
H&=&E+\frac{1}{2}\Gamma _{0}m_{\pi }^{2}+\frac{1}{2{\cal I}_{1}}\vec{\pi}_{h}^{2}
+\frac{1}{4{\cal I}_{2}^{\prime }}\pi _{s}^{\dag }\pi _{s}-i%
\frac{N}{8{\cal I}_{2}^{\prime }}(D^{\dag }\pi _{s}-\pi _{s}^{\dag }D) 
\nonumber  \label{hamil} \\
& &+\left[\frac{N^{2}}{16{\cal I}_{2}^{\prime }}+\Gamma _{0}(\chi
^{2}m_{K}^{2}-m_{\pi }^{2})+\Gamma_{3}\right] D^{\dag }D
+i\left[ \frac{1}{2{\cal I}_{1}}-\frac{1}{4{\cal I}_{2}^{\prime}}\left(1+\frac{\Gamma _{2}}{%
{\cal I}_{1}}\right) \right]  \nonumber \\
& &\cdot (D^{\dag}\vec{\pi}_{h}\cdot\vec{\tau}\pi _{s}-\pi _{s}^{\dag }\vec{\pi}_{h}\cdot\vec{\tau}D)
+\frac{N}{4{\cal I}_{2}^{\prime }}\left(1+\frac{\Gamma _{2}}{{\cal I}_{1}}%
\right) D^{\dag }\vec{\pi}_{h}\cdot \vec{\tau}D  \nonumber \\
& &+\left[ \frac{1}{2{\cal I}_{1}}-\frac{1}{3{\cal I}_{2}^{\prime }}\left( 1+%
\frac{3}{2}\frac{\Gamma _{2}}{{\cal I}_{1}}\right) +\frac{\Gamma _{2}^{2}+%
{\cal I}_{1}(\Gamma _{1}-\Gamma _{2})}{8{\cal I}_{1}{\cal I}_{2}^{\prime 2}}%
\right] (D^{\dag }D)(\pi _{s}^{\dag }\pi _{s})  \nonumber \\
& &+\left[ \frac{1}{12{\cal I}_{2}^{\prime }}\left( 1+\frac{3}{2}\frac{%
\Gamma _{2}}{{\cal I}_{1}}\right) -\frac{1}{8{\cal I}_{1}}-\frac{\Gamma
_{2}^{2}-{\cal I}_{1}(\Gamma _{1}-\Gamma _{2})}{32{\cal I}_{1}{\cal I}%
_{2}^{\prime 2}}\right] (D^{\dag }\pi _{s}+\pi _{s}^{\dag }D)^{2}  \nonumber
\\
& &-\left( \frac{1}{8{\cal I}_{2}^{\prime }}+\frac{\Gamma _{1}-\Gamma _{2}}{%
32{\cal I}_{2}^{\prime 2}}\right) (D^{\dag }\pi _{s}-\pi _{s}^{\dag }D)^{2} 
\nonumber \\
& &-i\frac{N}{8}\left[ \frac{1}{{\cal I}_{2}^{\prime }}\left( 1-\frac{\Gamma
_{2}}{{\cal I}_{1}}\right) +\frac{\Gamma _{2}^{2}+2{\cal I}_{1}(\Gamma
_{1}-\Gamma _{2})}{2{\cal I}_{1}{\cal I}_{2}^{\prime 2}}\right] (D^{\dag
}\pi _{s}-\pi _{s}^{\dag }D)(D^{\dag }D)  \nonumber \\
& &+\left[\frac{N^{2}}{12{\cal I}_{2}^{\prime }}-\frac{2}{3}\Gamma _{0}(\chi
^{2}m_{K}^{2}-m_{\pi }^{2})-\frac{2}{3}\Gamma _{3}\right.  \nonumber \\
& &\left.+\frac{N^{2}}{32}\frac{\Gamma _{2}^{2}+2{\cal I}_{1} (\Gamma
_{1}-\Gamma_{2})}{{\cal I}_{1}{\cal I}_{2}^{\prime 2}}\right] (D^{\dag
}D)^{2},
\label{ham00}
\end{eqnarray}
where ${\cal I}_{2}^{\prime}={\cal I}_{2}+\frac{1}{4}\Gamma_{1}$.

Through the symmetrization procedure~\cite{hkp98}, we can obtain the Hamiltonian 
of the form 
\begin{eqnarray}
H &=&E+\frac{1}{2}\Gamma _{0}m_{\pi }^{2}+\frac{1}{2{\cal I}_{1}}(%
\vec{I}^{2}+\frac{1}{4})+\frac{1}{4{\cal I}_{2}^{\prime }}\pi _{s}^{\dag
}\pi _{s}-i\frac{N}{8{\cal I}_{2}^{\prime }}(D^{\dag }\pi _{s}-\pi
_{s}^{\dag }D)  \nonumber \\
& &+\left[ \frac{N^{2}}{16{\cal I}_{2}^{\prime }}+\Gamma _{0}(\chi
^{2}m_{K}^{2}-m_{\pi }^{2})+\Gamma_{3}\right] D^{\dag }D  \nonumber \\
& &+i\left[ \frac{1}{2{\cal I}_{1}}-\frac{1}{4{\cal I}_{2}^{\prime }}\left(
1+\frac{\Gamma _{2}}{{\cal I}_{1}}\right) \right] (D^{\dag }\vec{I}\cdot 
\vec{\tau}\pi _{s}-\pi _{s}^{\dag }\vec{I}\cdot \vec{\tau}D)   \nonumber\\
& &+\frac{N}{4{\cal I}_{2}^{\prime }}\left( 1+\frac{\Gamma _{2}}{{\cal I}_{1}%
}\right) D^{\dag }\vec{I}\cdot \vec{\tau}D+\cdots.
\label{nht}
\end{eqnarray}
where the isospin operator $\vec{I}$ is given by $\vec{I}=\vec{\pi}_{h}$ and the ellipsis 
stands for the strange-strange interaction terms of order $1/N$ which can be readily read 
off from Eq. (\ref{ham00}).  Here one notes that the overall energy shift $\frac{1}{8{\cal I}_{1}}$
originates from the Weyl ordering correction in the BFT Hamiltonian scheme.  (See Ref.~\cite{hp00} 
for details.) 

Following the quantization scheme of Klebanov and Westerberg for the
strangeness flavor direction~\cite{kleb94}, one can obtain the Hamiltonian
of the form 
\begin{eqnarray}
H&=&E+\frac{1}{2}\Gamma _{0}m_{\pi }^{2}+\frac{1}{2{\cal I}_{1}}(%
\vec{I}^{2}+\frac{1}{4})+\frac{N}{8{\cal I}_{2}^{\prime }}(\mu -1)a^{\dag }a
\nonumber \\
&&+\left[ \frac{1}{2{\cal I}_{1}}-\frac{1}{4{\cal I}_{2}^{\prime }\mu }%
\left( 1+\frac{\Gamma _{2}}{{\cal I}_{1}}\right) (\mu -1)\right] a^{\dag }%
\vec{I}\cdot \vec{\tau}a  \nonumber \\
&&+\left[\frac{1}{8{\cal I}_{1}}-\frac{1}{8{\cal I}_{2}^{\prime}\mu ^{2}}
\left(1+\frac{\Gamma _{2}}{{\cal I}_{1}}\mu \right.\right.  \nonumber \\
&&\left.\left.-\frac{\Gamma _{2}^{2}+2{\cal I}_{1} (\Gamma _{1}-\Gamma _{2})%
}{4{\cal I}_{1}{\cal I}_{2}^{\prime}}(\mu -1)\right) (\mu -1)\right](a^{\dag
}a)^{2},   \label{htilde3}
\end{eqnarray}
where 
\begin{eqnarray}
\mu &=&\left( 1+\frac{\chi ^{2}m_{K}^{2}-m_{\pi }^{2}+\Gamma _{3}/\Gamma _{0}%
}{m_{0}^{2}}\right) ^{1/2},  \nonumber \\
m_{0} &=&\frac{N}{4(\Gamma _{0}{\cal I}_{2}^{\prime })^{1/2}}  \nonumber
\end{eqnarray}
and $a^{\dag }$ is creation operator for constituent strange quarks and we
have ignored the irrelevant creation operator $b^{\dag}$ for strange
antiquarks~\cite{kleb94}. Then, introducing the angular momentum of the strange
quarks 
\[
\vec{J}_{s}=\frac{1}{2}a^{\dag }\vec{\tau}a, 
\]
one can rewrite the Hamiltonian (\ref{htilde3}) as 
\begin{equation}
H=E+\frac{1}{2}\Gamma _{0}m_{\pi }^{2}+\omega a^{\dag }a+\frac{1}{2%
{\cal I}_{1}}\left( \vec{I}^{2}+2c\vec{I}\cdot \vec{J}_{s}+\bar{c}\vec{J}%
_{s}^{2}+\frac{1}{4}\right)  \label{hjs}
\end{equation}
where 
\begin{eqnarray}
\omega &=&\frac{N}{8{\cal I}_{2}^{\prime }}(\mu -1),  \nonumber \\
c&=&1-\frac{{\cal I}_{1}}{2{\cal I}_{2}^{\prime }\mu }\left( 1+\frac{\Gamma
_{2}}{{\cal I}_{1}}\right) (\mu -1),  \nonumber \\
\bar{c}&=&1-\frac{{\cal I}_{1}}{{\cal I}_{2}^{\prime }\mu ^{2}}\left( 1+%
\frac{\Gamma _{2}}{{\cal I}_{1}}\mu -\frac{\Gamma _{2}^{2}+2{\cal I}%
_{1}(\Gamma _{1}-\Gamma _{2})}{4{\cal I}_{1}{\cal I}_{2}^{\prime }}(\mu
-1)\right) (\mu -1).  \nonumber
\end{eqnarray}
Here note that the FSB effects are included in $c$ and $\bar{c}$, through 
$\Gamma_{1}$, $\Gamma_{2}$, ${\cal I}_{2}^{\prime}$ and $\chi$ and 
$\Gamma_{3}$ in $\mu$.

The Hamiltonian (\ref{hjs}) then yields the structure of the hyperfine
splittings as follows 
\begin{eqnarray}
\delta M &=&\frac{1}{2{\cal I}_{1}}\left[cJ(J+1)+(1-c) \left( I(I+1)-\frac{%
Y^{2}-1}{4}\right)\right.  \nonumber \\
&&\left.+(1+\bar{c}-2c)\frac{Y^{2}-1}{4}+\frac{1}{4}(1+\bar{c}-c)\right],
\end{eqnarray}
where $\vec{J}=\vec{I}+\vec{J}_{s}$ is the total angular momentum of the
quarks, and $c$ and $\bar{c}$ are the modified quantities due to the
existence of the FSB effect as shown above.

Now using the experimental values of the pion and kaon decay constants 
$f_{\pi}=93$ MeV and $f_{K}=114$ MeV, we fix the value of the Skyrmion 
parameter $e$ to fit the experimental data of $c_{exp}=0.67$ to 
yield the predictions for the values of $c$ and $\bar{c}$  
\begin{equation}
c=0.67,~~~\bar{c}=0.56
\end{equation}
which are contained in Table 1, together with the experimental data and the 
SU(3) rigid rotator predictions without pion mass.  For the massless and 
massive rigid rotator approaches we have used the above values for the decay constants $f_{\pi}$ and $f_{K}$ to obtain both the predictions in the FS and 
FSB cases.  As a result, we have explicitly shown that the more realistic physics considerations via the pion mass and the FSB terms improve both the 
$c$ and $\bar{c}$ values, as shown in Table 1.

\vskip 0.7cm \noindent 
{\Large {\bf Acknowledgments}}
\vskip 0.7cm \noindent 
STH would like to thank G.E. Brown for constant concerns and encouragements.  
YJP acknowledges financial support in part from the Korean Ministry of Education, 
BK21 Project No. D-1099.  The work of STH is supported in part by Grant No. 
2000-2-11100-002-5 from the Basic Research Program of the Korea Science and 
Engineering Foundation.

\begin{table}[h]
\caption{The values of $c$ and $\bar{c}$ in the massless pion and massive pion 
rigid rotator approaches to the SU(3) Skyrmions compared with experimental 
data. For the rigid rotator approaches, both the predictions in the flavor 
symmetric (FS) case and flavor symmetry breaking (FSB) one are listed.}
\begin{center}
\begin{tabular}{lcc}
\hline
Source & $c$ & $\bar{c}$ \\ \hline
Rigid rotator, massless and FS  & 0.92 & 0.86 \\ 
Rigid rotator, massless and FSB & 0.82 & 0.69 \\ 
Rigid rotator, massive and FS   & 0.79 & 0.66 \\ 
Rigid rotator, massive and FSB  & 0.67 & 0.56 \\ 
Experiment & 0.67 & 0.27 \\ \hline
\end{tabular}
\end{center}
\end{table}


\begin{thebibliography}{99}
\bibitem{adkins83}  G. S. Adkins, C. R. Nappa, and E. Witten, NaCl. Whys.
B228, 552 (1983).

\bibitem{sk}  M. Rho, A. Goldhaber, and G.E. Brown, Phys. Rev. Lett. 51, 747
(1983).

\bibitem{hsk}  I. Zahed and G.E. Brown, Phys. Rep. 142, 1 (1986); S.-T.
Hong, Phys. Lett. B417, 211 (1998).

\bibitem{su3}  E. Witten, Nucl. Phys. Rev. B223, 422 (1983); 
               E. Witten, Nucl. Phys. Rev. B223, 433 (1983);
               P.O. Mazur, M.A. Nowak, and M. Praszalowicz, Phys. Lett. B147, 137 (1984);
               SU(3) Skyrmion, Skyrmions and Anomalies, eds. M. Jezabek and M. Praszalowicz 
               (World Scientific, 1987) and references therein.

\bibitem{fsbsu3}  S.-T. Hong and B.-Y. Park, Nucl. Phys. A561, 525 (1993); 
                  S.-T. Hong and G.E. Brown, Nucl.Phys. A564, 491 (1993);
                  S.-T. Hong and G.E. Brown, Nucl. Phys. A580, 408 (1994).

\bibitem{yabu}  H. Yabu and K. Ando, Nucl. Phys. Rev. B301, 601 (1988).

\bibitem{multi}  J.H. Kim, C.H. Lee, and H.K. Lee, Nucl. Phys. A501, 835
(1989); H.K. Lee and D.-P. Min, Phys. Lett. B219, 1 (1989).

\bibitem{callan}  C.G. Callan and I. Klebanov, Nucl. Phys. B262, 365 (1985);
.

\bibitem{sco}  N.N. Scoccola, H. Nadeau, M.A. Nowak and M. Rho, Phys. Lett.
B201, 425 (1988).

\bibitem{di}  P.A.M. Dirac, Lectures in Quantum Mechanics (Yeshiva
University, New York, 1964).

\bibitem{BFT}  I. A. Batalin and E. S. Fradkin, Phys. Lett. B180, 157
(1986); Nucl. Phys. B279, 514 (1987); I. A. Batalin and I. V. Tyutin, Int.
J. Mod. Phys. A6, 3255 (1991).

\bibitem{BFT1}  R. Banerjee, Phys. Rev. D48 (1993) R5467; W.T. Kim, Y.-J.
Park, Phys. Lett. B336 (1994) 376.

\bibitem{kpr}  Y.-W. Kim, Y.-J. Park, K.D. Rothe, J. Phys. G24 (1998) 953;
Y.-W. Kim, K. D. Rothe, Nucl. Phys. B510 (1998) 511; M.-I. Park, Y.-J. Park,
Int. J. Mod. Phys. A13 (1998) 2179.

\bibitem{hkp98}  W. Oliveira and J.A. Neto, Int. J. Mod. Phys. A12, 4895 (1997);
S.-T. Hong, Y.-W. Kim, and Y.-J. Park, Phys. Rev. D59, 114026 (1999).

\bibitem{hong00}  S.-T. Hong, Y.T. Kim, Y.-J. Park, and M.S. Yoon, Phys. Rev. D62, 
085010 (2000).

\bibitem{kleb94}  K.M. Westerberg and I.R. Klebanov, Phys. Rev. D50, 5834
(1994); I.R. Klebanov and K.M. Westerberg, Phys. Rev. D53, 2804 (1996).

\bibitem{pari91}  G. Pari, B. Schwesinger and H. Walliser, Phys. Lett. B255, 1 (1991).

\bibitem{hong93}  S.-T. Hong and B.Y. Park, Nucl. Phys. A561, 525 (1993).

\bibitem{wzw}  I. Wess and B. Zumino, Phys. Lett. B37, 95 (1971); E. Witten,
Nucl. Phys. B223, 422 (1983); ibid, 433 (1983).

\bibitem{kleb90}  D. Kaplan and I.R. Klebanov, Nucl. Phys. B335, 45 (1990).

\bibitem{schw92}  B. Schwesinger, Nucl. Phys. A537, 253 (1992).

\bibitem{schwei91}  B. Schwesinger and H. Weigel, Phys. Lett. B267, 438 (1991); 
B. Schwesinger and H. Weigel, Nucl. Phys. A540, 461 (1992).

\bibitem{hp00}  S.-T. Hong and Y.-J. Park, Mod. Phys. Lett. A15, 913 (2000).

\end{thebibliography}
\end{document}